# A Scientific Trigger Unit for Space-Based Real-Time Gamma Ray Burst Detection

# II - Data Processing Model and Benchmarks

Hervé Le Provost[1], Stéphane Schanne[2], Christophe Flouzat[1], Pierre Kestener[3], Thomas Chaminade[1], Modeste Donati[2], Frédéric Château[1], François Daly[2], Jean Fontignie[2]

*Abstract*–The Scientific Trigger Unit (UTS) is a satellite equipment designed to detect Gamma Ray Bursts (GRBs) observed by the onboard 6400 pixels camera ECLAIRs. It is foreseen to equip the low-Earth orbit French-Chinese satellite SVOM and acts as the GRB trigger unit for the mission. The UTS analyses in real-time and in great details the onboard camera data in order to select the GRBs, to trigger a spacecraft slew re-centering each GRB for the narrow field-of-view instruments, and to alert the ground telescope network for GRB follow-up observations. A few GRBs per week are expected to be observed by the camera; the UTS targets a close to 100% trigger efficiency, while being selective enough to avoid fake alerts. This is achieved by running the complex scientific algorithms on a radiation tolerant hardware, based on a FPGA data pre-processor and a CPU with a Real-Time Operating System. The UTS is a scientific software, firmware and hardware co-development. A Data Processing Model (DPM) has been developed to fully validate all the technical choices deeply impacted by the ITAR restriction applied to the development. The DPM permits to evaluate the processing power and the memory bandwidth, and to adjust the balance load between software and firmware. This paper presents the UTS DPM functionalities and architecture. It highlights the results obtained with the full GRB trigger algorithms implemented on a rad-tolerant ITAR-free processor.

## I. INTRODUCTION

The Scientific Trigger Unit (UTS) analyses the data coming from a 6400 pixels array to detect in real-time all Gamma Ray Bursts (GRBs) for the SVOM space mission, and to inform the ground and the spacecraft of a GRB. The UTS continuously looks for an increase in the detector count-rates, in which case a sky image analysis is launched. It also periodically performs a full sky image analysis independently of any prior detector count-rate increase. In those reconstructed sky images, it searches for the presence of a new unknown gamma-ray source, called a GRB candidate. It sends its position to ground and to the spacecraft, which autonomously reorients itself to place the GRB candidate into the fields of view of the onboard narrow field instruments. These real-time tasks of the UTS involve complex algorithms which can only be performed by processors. To ease its processing load, dedicated FPGA logic devices are used to pre- and post-process the data analyzed by the processor.

## II. FUNCTIONALITIES

The UTS system (Fig. 1) is built around a single AT697F processor [1] in charge of running the GRB trigger algorithms. The observed GRB phenomena are expected to spread over a scale range varying from a few 10 ms to about 1300 s with various shapes in light curves. The best real-time GRB detection method would be to perform a full sky image analysis for all the time-scales and typically 4 energy bands. Because of the limited AT697F processing power and memory bandwidth, one full sky image analysis for one energy band typically takes 2 s. Therefore, for time-scales smaller than typically 20 s, a full sky image analysis will be triggered only upon a prior detection of an increase in the counting rate; this is called the "count-rate trigger". In parallel, to detect slow GRB phenomena, on time-scales larger than typically 20 s, a sky image analysis will be performed typically every time-window of 20 s and larger time-scales are built by summation of these time-windows; this is called the "image trigger".

Both trigger types are running on the same AT697F processor. A dedicated FPGA logic device is coupled to the processor via a PCI bus. It reads out the pixels data and performs on the fly some processing in order to lighten the processor load. The processed data are continuously written into a portion of the central SDRAM memory of the processor board. Every typically 10 ms the logic device freezes the data writing before emitting an interrupt to the processor. Each interrupt service routine retrieves from the FPGA the number of data written into the SDRAM since the previous interrupt. Once interrupts are served, the data writing process goes on.

The system is under control of the onboard payload computer (F-CU) via a SpaceWire link. The UTS system runs a state machine whose transitions are ordered by the F-CU. It is first configured with all the parameters related to the next acquisition phase, and then switched into its nominal acquisition mode. The firmware data acquisition/processing and the software GRB trigger algorithms are only performed in acquisition mode. A VHF alert message is posted to the SpaceWire link upon detection of a GRB. A second dedicated FPGA logic device, mounted on the processor board and coupled to the processor on its memory bus, is used to manage the SpaceWire link. This device formats and sends over the

SpaceWire the camera raw data. It also embeds a LICE interface used to observe and monitor without intrusion the real-time software ran by the AT697F processor.

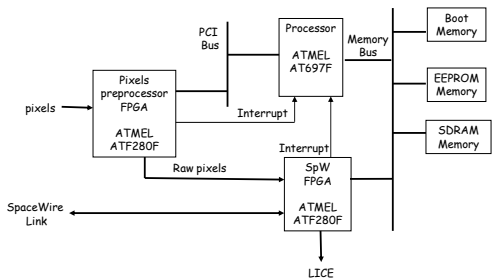

Fig. 1. Scheme of UTS hardware.

## III. UTS Hardware

Since the UTS is foreseen to equip a satellite built in cooperation with China, the UTS is built using exclusively ITAR-free components. It is a major constraint which prevents the UTS design from implementing the latest US space-technology components.

The selected processor is the ATMEL AT697F, packaged in a 256 pins MQFP. It is a 32-bit radiation-hardened SPARC processor delivering 86 MIPS and 23 MFLOPS at its maximum 100 MHz operating frequency. It integrates a 32-bit PCI core running at 33 MHz. The PCI interface is used to connect the preprocessor FPGA logic device. At power up, the logic device acts only as a PCI target and is initialized by the processor via the PCI. Once initialized and set into acquisition mode, it acts as a PCI master to write the processed data on the PCI bus. Those PCI cycles are then converted by the processor core into memory cycles on the dedicated processor memory bus. The effective input photon-rate sustained by the UTS has been measured on target with a PCI clock running at 20 MHz. It reaches 214000 photons/s compared to the 4000 photons/s nominal detector counting rate.

The selected logic device is the ATF280F [2] packaged in a 352 pins MQFP. It is a radiation hardened SRAM-based FPGA. All its cells are SEE-hardened and there is no need for TMR or SRAM configuration scrubbing. The ATF280F presents 231 I/O which are PCI compatible. It can be directly connected to the AT697F processor PCI port. It also embeds 115 kb of distributed RAM, mainly used to buffer the SpaceWire commands and buffer the pixels data at different processing stages.

The targeted processor core and external memory bus frequency is 100 MHz. The technological limitations of the ATF280F device and its associated development tools limit the targeted PCI system frequency. It is targeted to be in the range 20 to 33 MHz and will be fixed according to the final firmware design reports. The processor system hosts 128 kB of EEPROM used as PROM, 4 MB of EEPROM and 512 MB of SDRAM. The PROM holds the processor power-up boot software. It implements the vital hardware sanity-check functions and answers to a reduced set of SpaceWire commands. It copies the EEPROM software into the SDRAM and executes it upon reception of a specific command. The EEPROM holds the full scientific software. Once copied and executed in SDRAM, it runs a Real-Time Operating System (RTOS). The RTOS manages the tasks performing the GRB trigger and the tasks controlling the SpaceWire link. The EDAC integrated in the AT697F detects any double error and corrects any single error on a 32-bit bus. The SDRAM is sensitive to SEU and a multiple bit-flip in a critical single 32-bit word may corrupt the UTS behavior. Projected analysis on SEE effect over the SDRAM indicates that the usage of the AT697F integrated EDAC alone will lead to a reboot of the processor at a frequency acceptable for the ECLAIRs instrument.

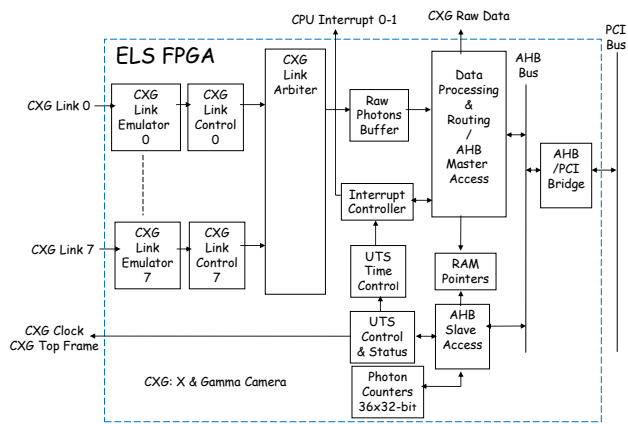

Fig. 2. Scheme of UTS firmware.

A simplified view of the firmware implemented in the ATF280F is shown Fig. 2. The design is fully synchronous to the PCI clock. It implements a PCI/AHB bridge. The device is initialized by the processor board through PCI master cycles decoded by the firmware as AHB slave accesses. The processor notably initializes the embedded RAM pointer area with the SDRAM pointers where should be written the different camera processed data. It then starts the camera serial-links data acquisition. A photon event ends in the raw-photons buffer, it is then readout and processed. Each processed data looks up for its associated SDRAM pointer in the RAM pointers area. It then triggers an AHB master cycle resulting in an AT697F memory cycle. The firmware internally keeps updated the different word counts of processed data written into the SDRAM. Every typically 10 ms, it freezes the raw-photon buffers readout, interrupts the processor. The interrupt subroutine reads the word counts and unfreezes the photon-buffers readout. The pixel counting buffers (called Shadowgrams) are located in the SDRAM memory and accessed through PCI Read/Modify/Write cycles. The ATF280F device occupation rate is 65% with 30% of the RAM blocks used. Another critical point is the maximum system operating frequency. It is currently limited to 21 MHz and the minimal targeted operating frequency is 20 MHz.

## IV. DATA PROCESSING MODEL

### A. Architecture

In order to test and validate the design of the UTS, a Data Processing Model (DPM) has been developed. The DPM implements the pixels preprocessing FPGA, connected to an AT697F processor via PCI dialogue. The DPM permits to validate the choice of share between firmware and software, as well as realistic benchmarks of the algorithms on the AT697F.

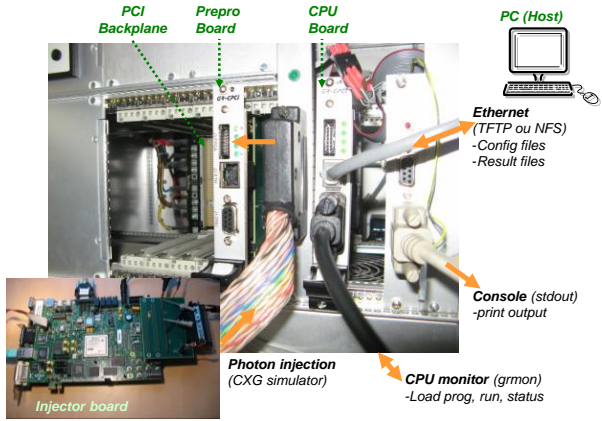

Fig. 3. DPM Hardware.

The DPM reproduces the UTS hardware architecture. It is built of commercial readily available boards assembled in a compact PCI crate (Fig. 3). The PCI setup is functionally representative of the camera data firmware and software processing. The DPM allows focusing on the UTS specific software and firmware developments. The AT697F processor is the only component fully representative of the UTS flight-model. The pixel preprocessing logic is implemented in a XC4VLX100 instead of the ATF280F. The DPM does not implement the SpaceWire link. It is connected to a PC station running LINUX through an Ethernet link and two UARTs. One UART is connected to the processor standard IO. It is used for basic debugging and real-time monitoring and permits to print onto a console. The second UART is connected to the processor monitor and managed by the GAISLER [3] GRMON monitor. It allows downloading programs on the AT697F target and controlling its execution. The Ethernet port is used to read the configuration files of the algorithms and to dump the processing results in files for offline analysis. A photon injector emulates the pixel links. It receives pre-calculated and time-stamped photon-files from the PC station. The PC station synchronizes the DPM elements. It first downloads the photon-file to the injector and initializes the AT697F processor board. Once the AT697F processor is in acquisition mode, it starts the photons injector. The dumped result-files can then be checked against the downloaded photon-files.

The processor is running the RTEMS RTOS [4]. RTEMS provides the multitasking feature required to run the image and count-rate trigger processes. The included network services and POSIX standards are key features. Coupled to the GAISLER tools, the RTEMS ecosystem allowed speeding up the development. The use of RTEMS compared to the Bare-C tool chain amounts to have executable files which are several 100 kB larger depending on which services are enabled.

### B. Development methodology

The AT697F processing power and the ATF280F logic capacities are limited compared to today's top commercial embedded processors and logic devices but they are the best available devices for our application considering the ITAR restriction. To realistically evaluate them for our application, we use the development methodology shown on Fig. 4. The GRB detection algorithms are developed on the Scientific Software Model (SSM) but the SSM performances in terms of CPU usage and memory bandwidth can be hardly extrapolated to the UTS architecture. The SSM is meant to simulate multiples orbits in a few minutes and to efficiently develop and tune the algorithms. On the contrary, the photons are injected on the DPM in real-time and a 2 hours sky simulation takes 2 hours to be processed by the DPM. The DPM realistically takes into account the pixels data preprocessing and data flow, the PCI and AT697F memory bandwidth, the processor computing power. The CPU source code used in the SSM and the DPM are updated with the latest algorithms, both simultaneously since they both link with their specific applications the same libraries implementing the scientific trigger algorithms.

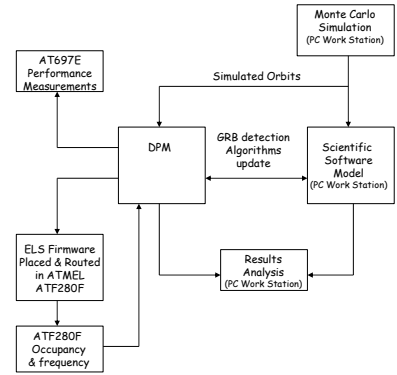

Fig. 4. DPM Development.

The firmware of the DPM is developed in VHDL and the source code is in parallel synthesized to produce a net list for the ATF280F part. The net list is then placed and routed using the ATMEL tools. It gives an occupancy rate and a maximum operating frequency. The VHDL code is developed taking into account the ATF280F restriction.

### C. FPGA driver

The scheme of the dialogue between FPGA and CPU is shown on Fig. 5. For every photon received on one of the 8

input links, the energy band (out of 8) and energy strip combination (out of 4) as well as the detector zone (out of 9) are computed. Each 32 bit-wide photon received on a link is stored into its associated raw data ring-memory. The raw data buffers are heavily used for system debugging and validation. To store 20 min of data, at an input rate of ~4000 photons/s for the whole camera and a margin factor of ~3.5 applied, each of the 8 rings is allocated with 8 MB. From the 32-bit wide raw data photons, the FPGA produces 16-bit wide preprocessed photons containing the pixel number out of 6400 and the energy band. These are stored in 1 photon data ring-memory. The photon classification performed by the FPGA generates 36 counters (number of photons per Energy Strip and Detector Zone combination), handed over each 10 ms to the CPU during the interrupt routine. These are stored into incremental counting ring-memories. The photon buffer and the counters are to be used by the count-rate trigger.

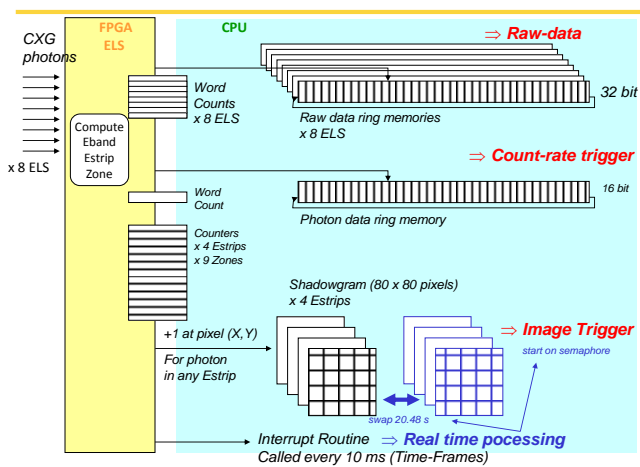

Fig. 5. Firmware pixels preprocessing

The FPGA fills 1 shadowgram per energy strip used by the image trigger. These Shadowgrams are implemented in the RAM of the CPU board in a double buffer mode: while the FGPA is filling the hidden layer with photons, the CPU is analyzing the visible layer. The switch between the layers is performed in the interrupt routine, every 20480 Time Frames (of 10 ms).

## V. DPM RESULTS

Currently the UTS scientific software has been implemented in the DPM and fully validated. It covers the FPGA driver, and the scientific algorithms. The count-rate trigger and image trigger algorithms, developed on the SSM, have been linked with the RTEMS OS application on the Leon2 CPU of the DPM. The Leon2 CPU is running at 100 MHz and the PCI clock is downscaled to 20 MHz. The detailed algorithms are described in [5].

### A. Image trigger benchmark

As an input we select randomly one half-orbit, with an Earth entry in the Field Of View (FOV). We add randomly three GRBs. The resulting photon file is processed by the SSM and DPM, the results of the processing is plotted on Fig. 6, showing as a function of time the time intervals resulting in the most significant excess found (in blue below threshold, in purple above threshold, i.e. GRB detection). DPM and SSM results do not show any significant difference in GRB detection.

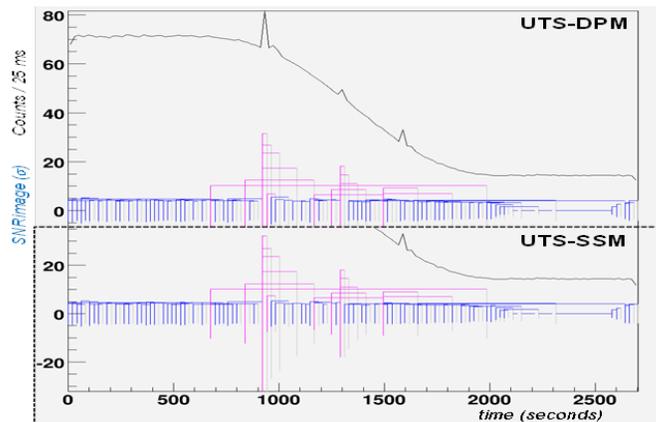

Fig. 6. Result of the image trigger on half an orbit with Earth entry into FOV, including three GRBs. Top: data processed by DPM. Bottom: data processed by SSM.

The execution time of the image trigger function over the simulated half-orbit (45 min, 130 steps of 20.48 s) is presented in Fig. 7. The execution requires at least 6 s, with peaks up to 9 s in the case of GRB detection. The processing time depends only very slightly on the data volume which is normal, since the algorithm works on pre-computed shadowgrams. Most of the variability in the processing time is due to the structure of the time-history algorithm (for which in special phases of the algorithm all time scales are evaluated together).

In details here are the computation times of the different algorithm parts:

- Sky exposure: 132 ms mean, 115 ms to 200 ms peak, according to time history phase. This time is consumed only once in the 20.48 s period.

All times below are consumed 4 times (one per each energy strip):

- Spatial background fit on shadowgram: 577 ms mean, 567 ms to 580 ms peak.

- Deconvolution including counts and variance: 840 ms mean.

- Sky summation: 55 ms mean, 30 ms to 180 ms peak, according to time history phase.

- Excess search: 105 ms mean, 40 ms to 300 ms peak, according to time history phase.

- Excess position fit: 200 ms to 400 ms, in case of excess above threshold only.

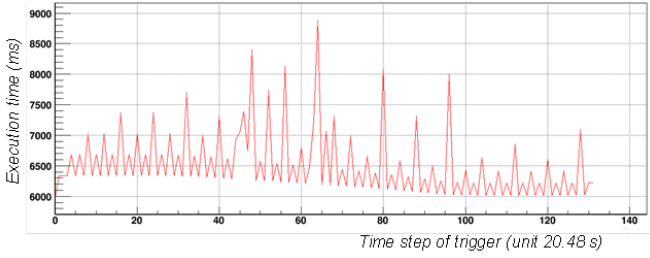

Fig. 7. Execution time of the image trigger by DPM as a function of time.

### B. Count-rate trigger benchmark

The Fig. 8 shows as a function of time the time intervals resulting in the most significant excess found (in blue below threshold, in purple above threshold, i.e. GRB detection). For the full detector zone and full energy band only, in black is shown the counts per 2.56 s time interval, and in green the corresponding background estimate over which the count-rate excesses are searched.

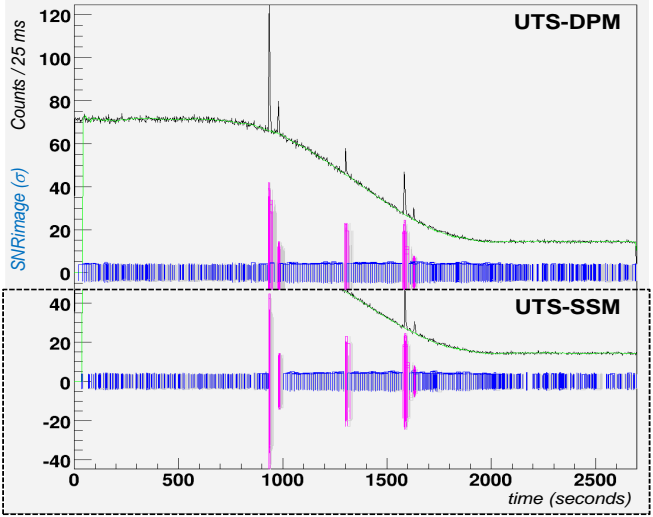

Fig. 8. Result of the count-rate trigger on half an orbit with Earth entry into FOV, including three GRBs. Top: data processed by DPM. Bottom: data processed by SSM.

The Fig. 9 shows a zoom on the first, 2nd and 3rd GRB detection by the count-rate trigger algorithm. Each 2.56 s (marked by a grey vertical line), using a smoothly declining background estimate, the algorithm determines excesses over this background estimate (first part of the algorithm), and performs the evaluation of the most significant excess not yet analyzed by deconvolution (second part of the algorithm). Deconvolved time windows (along x axis) and their significances (along y axis) are marked by blue boxes (excess in image below threshold) and purple boxes (excess above threshold, GRB found).

In this example the later in time, the less significant are the detections (boxes are less high at later moments of evaluation, marked by the grey vertical line).

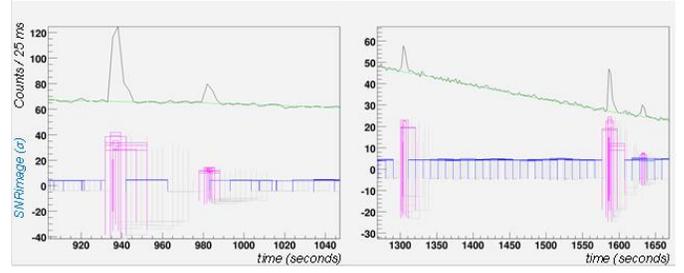

Fig. 9. Zoom on the first GRB (left) and second and third GRB (right) processed by DPM.

The execution time of the count-rate trigger function over the simulated half-orbit (45 min, 1050 steps of 2.56 s) is presented in Fig. 10. As we can see, the execution requires at least 300 ms when no count-rate excess is deconvolved, and about 1.3 s with count-rate excess deconvolution, and it peaks up to 1.6 s when a significant excess is found in the image and is fitted.

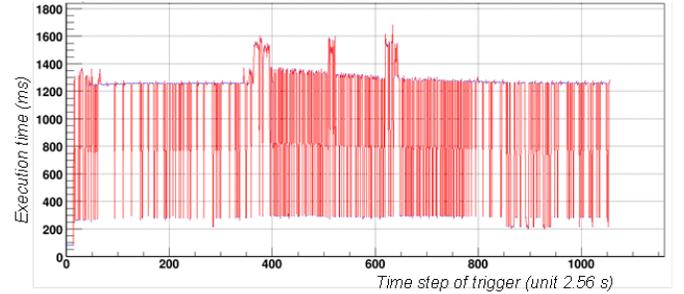

Fig. 10. Execution time of the count-rate trigger by UTS-DPM as a function of time.

In details, here are the computation times of the different algorithm parts:
- Temporal background estimate on all 9 zones and 4 energy strips: 60 ms – 70 ms.
- Search for count-rate excesses on all 12 time scales from 10 ms to 20.48 s: 100 ms – 130 ms.
- Extraction of best excess from buffer: 70 ms
- Localization of best excess: 1 s – 1.3 s in case of excess position fit.

### C. Benchmark summary

The Benchmarks of both scientific triggering algorithms completely implemented on the DPM can be summarized as follows:
- The image trigger task, executed every 20.48 s, runs in 6 s to 9 s (mean: 6.51 s, i.e. 31.8% of the time slot).
- The count-rate trigger task, executed every 2.56 s, runs in 0.3 s to 1.6 s (mean: 1.01 s, i.e. 39.4% of the time slot).

Based on these first measurements of the performances of both scientific trigger algorithms implemented, we can conclude that one Leon2-CPU at 100 MHz offers sufficient CPU power to achieve the UTS scientific tasks. For the other management tasks, the currently estimated margin available is 28% of the CPU power.

Even if the count-rate trigger algorithms were tuned in such a way that one deconvolution is performed each time-slot of 2.56 s, the mean execution time would only raise to about 1.4 s, i.e. 55% of the time slot. In this case using one Leon2-CPU at 100 MHz the remaining CPU power margin would be of 13% which is still acceptable.

ACKNOWLEDGMENT

The technical developments presented in this paper are co-financed by the CEA and the French Space Agency CNES.